\input epsf
\documentstyle{mn}
\newif\ifAMStwofonts

\def\lesssim{\mathrel{\hbox{\rlap{\hbox{\lower4pt\hbox{$\sim$}}}\hbox{$<$}}}}
\def\gtrsim{\mathrel{\hbox{\rlap{\hbox{\lower4pt\hbox{$\sim$}}}\hbox{$>$}}}}
\def\apj{ApJ}

\def\aj{AJ}
\def\aap{A\&\hskip-1pt A}

\def\mnras{MNRAS}

\def\eqalign#1{\null\,\vcenter{\openup\jot
  \ialign{\strut\hfill$\displaystyle{##}$&$
      \displaystyle{{}##}$\hfill \crcr#1\crcr}}\,}

\newcommand{\Deltavec}{\mbox{\boldmath $\Delta$}}
\newcommand{\deltavec}{\mbox{\boldmath $\delta$}}
\newcommand{\thetavec}{\mbox{\boldmath $\theta$}}
\newcommand{\muvec}{\mbox{\boldmath $\mu$}}
\newcommand{\uvec}{\mbox{\boldmath $u$}}
\newcommand{\vvec}{\mbox{\boldmath $v$}}

\title[Parallax-induced Deviation]
      {Detectability of the Parallax-induced Deviations in the \\ 
       Astrometric Centroid Shift Trajectories of Gravitational \\
       Microlensing Events}
\author[Han \& Chang]
       {Cheongho Han, \& Kyongae Chang\\
       	Dept.\ of Astronomy \& Space Science, \\
        Chungbuk National University, Chongju, Korea 361-763\\
	{\tt cheongho@astro.chungbuk.ac.kr}\\
	 \\
	Department of Physics, \\
        Chongju University, Chongju, Korea 360-764 \\
        {\tt kchang@chongju.ac.kr}}

\date{Accepted:\\
      Received: }

\pagerange{\pageref{firstpage}--\pageref{lastpage}}

\begin{document}

\maketitle

\label{firstpage}

\begin{abstract}
The lens mass determined from the photometrically obtained Einstein time
scale suffers from large uncertainty due to the lens parameter degeneracy.
The uncertainty can be substantially reduced if the mass is determined
from the lens proper motion obtained from astrometric measurements of the
source image centroid shifts, $\deltavec\thetavec_{\rm c}$, by using high
precision interferometers from space-based platform such as the {\it Space
Interferometry Mission} (SIM) and ground-based interferometers soon available
on several 8 -- 10 m class telescopes.  However, for the complete resolution
of the lens parameter degeneracy it is required to determine the lens parallax
by measuring the parallax-induced deviations in the centroid shifts trajectory,
$\Deltavec\deltavec\thetavec_{\rm c}$.

In this paper, we investigate the detectabilities of $\deltavec
\thetavec_{\rm c}$ and $\Deltavec\deltavec \thetavec_{\rm c}$ by determining
the distributions of the maximum centroid shifts and the average maximum
deviations expected for different types of Galactic microlensing events
caused by various masses.  From this investigation, we find that as long as
source stars are bright enough for astrometric observations it is expected
that $\deltavec \thetavec_{\rm c}$ for most events caused by lenses with
masses greater than $0.1\ M_\odot$ regardless of the event types can be
easily detected from observations by using not only the SIM (with a detection
threshold $\delta\theta_{\rm th}\sim 3\ \mu{\rm as}$) but also the ground-based
interferometers (with $\delta \theta_{\rm th}\sim 30\ \mu{\rm as}$).  However,
detection of $\Deltavec \deltavec \thetavec_{\rm c}$ from ground-based
observations will be difficult for nearly all Galactic bulge self-lensing
events, and will be restricted only for small fractions of disk-bulge and
halo-LMC events, for which the deviations are relatively large.  From
observations by using the SIM, on the other hand, detecting $\Deltavec
\deltavec\thetavec_{\rm c}$ will be possible for majority of disk-bulge and
halo-LMC events and even for some fraction of bulge self-lensing events.  For
the complete resolution of the lens parameter degeneracy, therefore, SIM
observations (or equivalent) will be essential.
\end{abstract}


\section{Introduction}

The biggest difficulty in revealing the nature of Galactic dark matter by
using microlensing is that one cannot determine the individual lens masses
because the Einstein time scale, which is the only quantity related to the
lens mass obtained from the photometrically measured event light curve,
results from the combination of other lens parameters by
\begin{equation}
t_{\rm E}=\left[ 
{4GM\over c^2 v^2}{D_{ol}(D_{os}-D_{ol})\over D_{os}}
\right]^{1/2},
\end{equation}
where $M$ is the lens mass, $v$ is the lens-source transverse speed, and
$D_{ol}$ and $D_{os}$ represent respectively the distances to the lens and
the source from the observer.  As a method to resolve the lens parameter
degeneracy for general microlensing events, it was proposed to measure the
shifts of the source star image centroid caused by lensing from follow-up
astrometric observations for events detected photometrically from the ground
by using high precision interferometers from space-based platform such as the
{\it Space Interferometry Mission} (SIM) and ground-based interferometers
soon available on several 8 -- 10 m class telescopes such as the Keck and
the Very Large Telescope (Miyamoto \& Yoshii 1995; H\o\hskip-1pt g, Novikov,
\& Polarev 1995; Walker 1995; Dominik \& Sahu 2000).

When the lens-source separation is normalized by the angular Einstein ring
radius $\theta_{\rm E}$, the centroid shift vector with respect to the
position of the un-lensed source for an event caused by a point-mass lens is
represented by
\begin{equation}
\deltavec\thetavec_{\rm c} = {\uvec \over u^2+2} \theta_{\rm E};
\end{equation}
\begin{equation}
u_x = {t-t_0\over t_{\rm E}},\ u_y = u_{\rm min},
\end{equation}
where $\uvec$ is the lens-source separation normalized by $\theta_{\rm E}$,
$u_x$ and $u_y$ represent the components of $\uvec$ parallel and normal to
the lens proper motion vector $\muvec$, $t_0$ is the time of the closest
lens-source approach, and $u_{\rm min}$ is the lens-source separation (also
normalized by $\theta_{\rm E}$) at that moment (i.e.\ the impact parameter).
The angular Einstein ring radius is related to the physical parameters of
the lens by
\begin{equation}
\theta_{\rm E} = {vt_{\rm E}\over D_{ol}}
               = \left( {4GM\over c^2}{D_{os}-D_{ol}\over D_{ol}D_{os}}
                 \right)^{1/2}.
\end{equation}
When $u=\sqrt{2}$, the amount of the centroid shift becomes maximum with
\begin{equation}
\delta\theta_{\rm c,max}= {\theta_{\rm E}\over\sqrt{8}}.
\end{equation}
During the event, the trajectory of the centroid shifts traces an ellipse
(astrometric ellipse).  Once the trajectory is constructed from astrometric
observations of the event, one can determine $\theta_{\rm E}$ because the
size of the astrometric ellipse (semi-major axis) is directly proportional
to $\theta_{\rm E}$.  Then one can determine the lens proper motion by
$\mu=\theta_{\rm E}/t_{\rm E}$ combined with the photometrically determined
value of $t_{\rm E}$.  With the determined value of $\mu$, the uncertainty
in the determined lens mass can be substantially reduced because $\mu$ does
not depend on $v$.  However, the lens parameter degeneracy is not completely
broken even with the determination of $\mu$, because it still results from
the combination of the lens mass and the location.

The lens parameter degeneracy can be completely broken, and thus the lens
mass can be uniquely determined if one determines the lens parallax by
astrometrically measuring the deviations in the observed centroid shift
trajectory, $\Deltavec \deltavec\thetavec_{\rm c}$, caused by the change of
the observer's location during the event due to the orbital motion of the
Earth around the Sun: parallax effect.\footnote{The lens mass can also be
determined by astrometrically monitoring nearby very high proper motion stars
working as a lens instead of monitoring lensed source stars because the
distance to the lens is known.  This method was suggested by Miralda-Escud\'e
(1996) and Paczy\'nski (1998) and the specific candidate stars from the SIM
monitoring were selected by Salim \& Gould (2000).  We note, however, that
our analysis is focused on followup observations for events detected by the
photometric lensing surveys from the ground.} However, $\Deltavec\deltavec
\thetavec_{\rm c}$, in general, is much smaller than $\deltavec
\thetavec_{\rm c}$, and thus measurements of the parallax-induced deviations
require higher precision observations.  The possibility of astrometric
parallax measurements was investigated by Paczy\'nski (1998) and further
analysis considering the actual performance of the specific instruments was
done by Boden, Shao, \& Van Buren (1998).  However, their analyses were based
on several example events.

In this paper, we determine the distributions of the maximum centroid shifts,
$f(\delta\theta_{\rm c, max})$, and the average maximum deviations,
$f(\langle\Delta\delta\theta_{\rm c,max}\rangle)$, for different types of
Galactic microlensing events caused by various masses.  From the analysis of
these distributions, we statistically investigate the detectabilities of
$\deltavec\thetavec_{\rm c}$ and $\Deltavec\deltavec\thetavec_{\rm c}$
expected from ground- and space-based observations.

\begin{figure}
\epsfysize=10cm
\centerline{\epsfbox{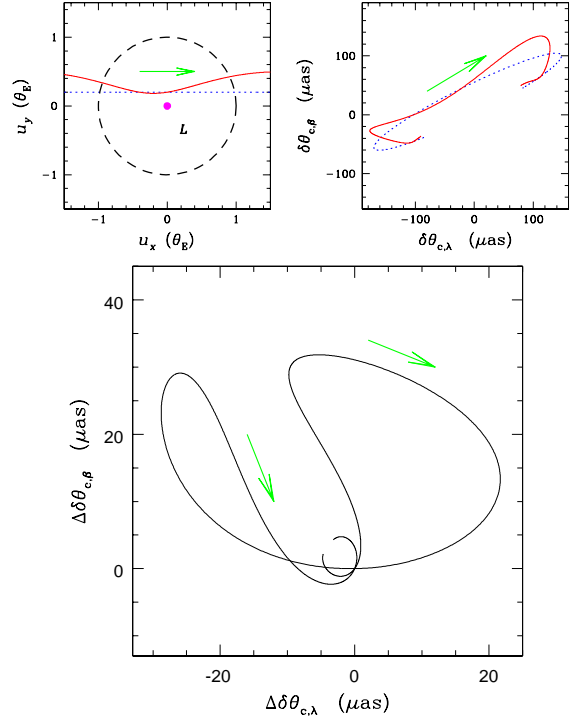}} 
\caption{
The parallax effect on the astrometric behavior of an example
halo-LMC event.  The upper left panel shows the source star trajectories
with respect to the lens ($L$) with (solid curve) and without (dotted curve)
the parallax effect.  The dashed circle centered at $L$ represents the
Einstein ring.  In the upper right panel, the trajectories of the centroid
shifts $\deltavec\thetavec_{\rm c}$ resulting from the individual source
trajectories in the upper left panel are marked by the same line types.
In the lower panel, we present the trajectory of the deviation vector
$\Deltavec\deltavec\thetavec_{\rm c}$, which corresponds to the vector
difference between the solid and dotted curve in the upper right panel.
Both the trajectories of $\deltavec\thetavec_{\rm c}$ and $\Deltavec\deltavec
\thetavec_{\rm c}$ are presented in the ecliptic coordinates.  For the
details about the lens system, see the text.
}
\end{figure}

\section{Parallax-induced Deviations in Centroid Shift Trajectories}

Due to the orbital motion of the Earth around the Sun, the lens location
in the ecliptic coordinates changes by
\begin{equation}
\Delta\varphi_\lambda = \Pi \sin(\lambda_\odot - \lambda),
\end{equation}
\begin{equation}
\Delta\varphi_\beta = -\Pi \sin\beta \cos(\lambda_\odot - \lambda),
\end{equation}
where $\Pi=1\ {\rm AU}/D_{ol}$ represents the lens parallax, $(\lambda,
\beta)$ is the ecliptic coordinates towards the direction where the event
is occurred, and $\lambda_\odot$ represents the ecliptic latitude of the
Sun during the event.  Then the change of the source location (normalized
by $\theta_{\rm E}$) with respect to the lens becomes
\begin{equation}
\Delta u_x = 
-\left( 
{\Delta\varphi_\lambda \over \theta_{\rm E}} \cos\phi + 
{\Delta\varphi_\beta \over \theta_{\rm E}} \sin\phi
\right)
\left( {D_{os}-D_{ol}\over D_{os}}\right),
\end{equation}
\begin{equation}
\Delta u_y = 
-\left( 
-{\Delta\varphi_\lambda \over \theta_{\rm E}} \sin\phi + 
{\Delta\varphi_\beta \over \theta_{\rm E}} \cos\phi
\right)
\left( {D_{os}-D_{ol}\over D_{os}}\right),
\end{equation}
where $\phi$ represents the angle between $\muvec$ and the ecliptic plane.
The factor $(D_{os}-D_{ol})/D_{os}$ is included because all lengths are
projected onto the source plane, and the minus sign is included because
the source position changes towards the direction opposite to that of the
lens position change.  Then the parallax-induced deviation vector of the
observed centroid shift from the centroid shift expected without the Earth's
orbital motion is obtained by
\begin{equation}
\Deltavec\deltavec\thetavec_{\rm c} = 
\deltavec\thetavec_{\rm c} (\uvec+\Deltavec\uvec) - 
\deltavec\thetavec_{\rm c} (\uvec),
\end{equation}
where $\Deltavec\uvec = (\Delta u_x,\Delta u_y)$.

The parallax effect on the astrometric behavior of a microlensing event
is demonstrated in Figure 1.  The upper left panel shows the source star
trajectories with respect to the lens ($L$) with (solid curve) and without
(dotted curve) the parallax effect.  In the upper right panel, the centroid
shift trajectories resulting from the individual source trajectories in
the upper left panel are marked by the same line types.  In the lower panel,
the trajectory of the deviation vector $\Deltavec\deltavec\thetavec_{\rm c}$
is presented, which corresponds to the vector difference between the solid
and dotted curves in the upper right panel.  The event is assumed to be
caused by a lens located in the Galactic halo with $D_{ol} =10\ {\rm kpc}$
and $M=0.5\ M_\odot$ for a source star located in the Large Magellanic
Cloud (LMC) with $D_{os}=50\ {\rm kpc}$.  We also assume that the lensing
parameters are $u_{\rm min}=0.2$ and $t_{\rm E}=100\ {\rm days}$, and the
closest lens-source approach occurs at the moment when the source is at
the largest angular separation from the Sun (i.e.\ $\lambda_\odot-\lambda
=180^\circ$).  Observations are assumed to be carried out during
$-5 t_{\rm E}\leq t_{\rm obs}\leq 5 t_{\rm E}$ with respected to the time
of maximum amplification.  The ecliptic latitude of a star in LMC is
$\beta\sim \sin^{-1}(-0.99)$.  Note that both the trajectories of
$\deltavec\thetavec_{\rm c}$ and $\Deltavec\deltavec\thetavec_{\rm c}$ are
presented in the ecliptic coordinates.  From the figure, one finds that
$\Delta\delta\theta_{\rm c}$ is much (nearly an order) smaller than
$\delta\theta_{\rm c}$.

\begin{figure}
\epsfysize=13cm
\centerline{\epsfbox{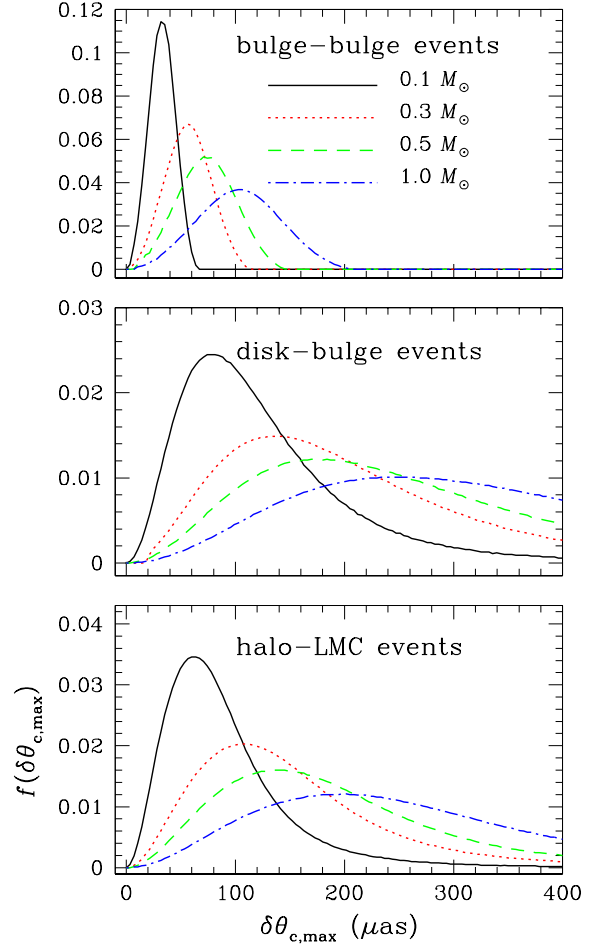}}
\caption{
The distributions of the maximum astrometric centroid shifts expected from
different types of Galactic events caused by various masses.  Note that the
distributions are arbitrarily normalized.
}
\end{figure}

\section{Detection of Centroid Shifts}

In this section, we show that as long as source stars are bright enough
for astrometric observations\footnote{Note that astrometric observations
of microlensing events are restricted by source brightness.  With the
SIM having a detection limit of $V\sim 20$ mag, although most Galactic
bulge events can be observed, a substantially fraction (nearly half) of
the LMC events cannot be observed owing to the faintness of their source
stars (Han \& Chang 1999).} one can detect $\deltavec\thetavec_{\rm c}$
for most Galactic events caused by lenses with masses greater than $0.1\ 
M_\odot$ regardless of the event types from observations by using not only
the SIM but also the ground interferometers.

To show this, we compute the distribution of the maximum centroid shifts by
\begin{equation}
\eqalign{
f(\delta\theta_{\rm c,max}) & = \int_0^\infty dD_{os} \rho(D_{os})
\int_0^{D_{os}} dD_{ol} \rho(D_{ol})\pi r_{\rm E}^2 \cr
 & \times \int_0^\infty \int_0^\infty dv_x dv_y v f(v_x,v_y) \cr
 & \times \delta \left[ \delta\theta_{\rm c,max} - 
  \theta_{\rm E}(D_{ol},D_{os},M)/\sqrt{8}\right], \cr 
}
\end{equation}
where $r_{\rm E}=D_{ol}\theta_{\rm E}$ is the physical radius of the
Einstein ring, $\rho(D_{ol})$ and $\rho(D_{os})$ represent the matter
density distributions of the lenses and source stars along the line of
sight, $\delta$ represents the dirac delta function, $v_x$ and $v_y$ are
the two components of $\vvec$, and $f(v_x,v_y)$ is their distribution.
For the physical and dynamical distributions of matter in the Galactic
disk and bulge, we adopt the models of Han \& Gould (1995).  For the
matter in the Galactic halo, we adopt an isothermal sphere model with
a core radius of the form
\begin{equation}
\rho(r) = \rho_0 {r_{\rm c}^2 + R_0^2 \over r_{\rm c}^2 + r^2},
\end{equation}
where $r$ is the distance measured from the Galactic center, $R_0=8\ 
{\rm kpc}$ is the adopted Galacto-centric distance of the Sun, and
$\rho_0=7.9\times 10^{-3}\ M_\odot\ {\rm pc}^{-3}$ is the halo mass
density in the solar neighborhood.  The adopted velocity dispersion and
the core radius are $\sigma=250/\sqrt{2}\ {\rm km}\ {\rm s}^{-1}$ and
$r_{\rm c} = 2\ {\rm kpc}$.  The factors $\pi r_{\rm E}^2$ and $v$ in
equation (11) are included to weight the event rate by the cross-section
(i.e.\ Einstein ring area) and the transverse speed.

\begin{figure}
\epsfysize=13cm
\centerline{\epsfbox{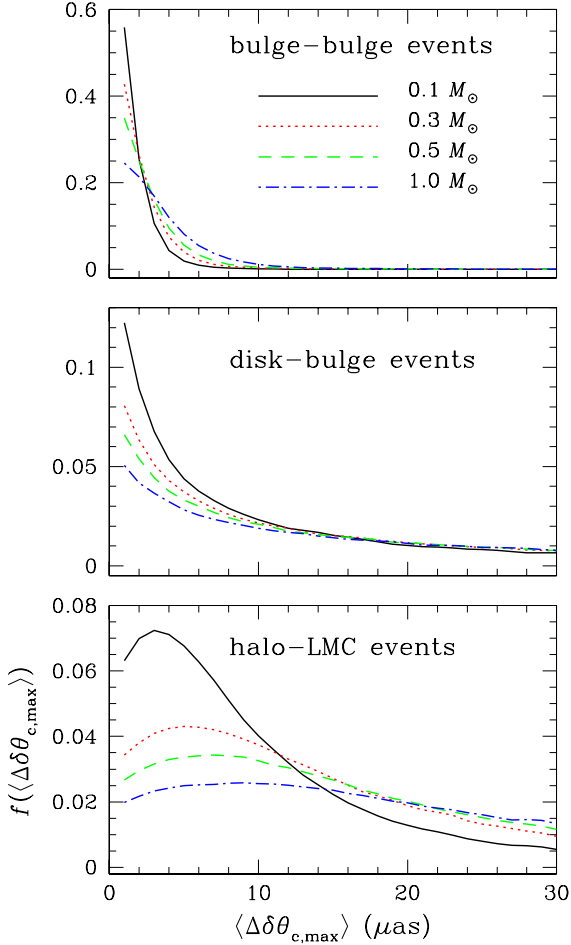}}
\caption{
The distributions of the average maximum deviations in the centroid shift
trajectory for different types of Galactic events caused by various masses.
Note that the distributions are arbitrarily normalized.
}
\end{figure}

In Figure 2, we present the computed distributions $f(\delta\theta_{\rm c, 
max})$ for different types of Galactic events.\footnote{We note that both
the distributions $f(\delta\theta_{\rm c,max})$ and $f(\langle\Delta\delta
\theta_{\rm c,max})\rangle$ are determined for all events with $u_{\rm min}
\leq 1.0$, not considering the dependence of the detection efficiency on
the event duration and the amplification.  As a result, they are different
from the distributions for events actually detected by the previous and the
current lensing surveys (Alcock et al.\ 1993; Aubourg et al.\ 1993; Udalski
et al.\ 1993; Alard \& Guibert 1995; Abe et al.\ 1997).  We determine the
distributions in this way because the detection efficiency is greatly
dependent on the adopted observational strategy, the instruments, and the
data analysis technique, which are evolving rapidly, and thus the analysis
based on the observation condition and strategy of a specific lensing survey
will have little meaning.}  In each panel, we mark the type of events
according to the order of the lens and source locations.  For example,
`disk-bulge events' refer to those caused by lenses located in the Galactic
disk for source stars located in the bulge.  From the obtained distributions,
one finds that for a given lens mass bulge self-lensing (bulge-bulge) events
are most likely to have the smallest value of $\delta\theta_{\rm c,max}$.
One also finds that the expected most frequent value of
$\delta\theta_{\rm c,max}$ even for this type of events caused by small
mass lenses with $M\sim 0.1\ M_\odot$ exceeds $\sim 30\ \mu{\rm as}$.
Then, considering the detection threshold is $\delta\theta_{\rm th}\sim 3\ 
\mu{\rm as}$ for the SIM (Unwin, Boden, \& Shao 1997) and $\delta
\theta_{\rm th}\sim 30\ \mu{\rm as}$ for the ground interferometers
(Colavita et al.\ 1998), detection of $\deltavec\thetavec_{\rm c}$ will
be possible for most of events regardless of the event types from
observations by using not only the SIM but also the ground interferometers.

\begin{figure}
\epsfysize=13cm 
\centerline{\epsfbox{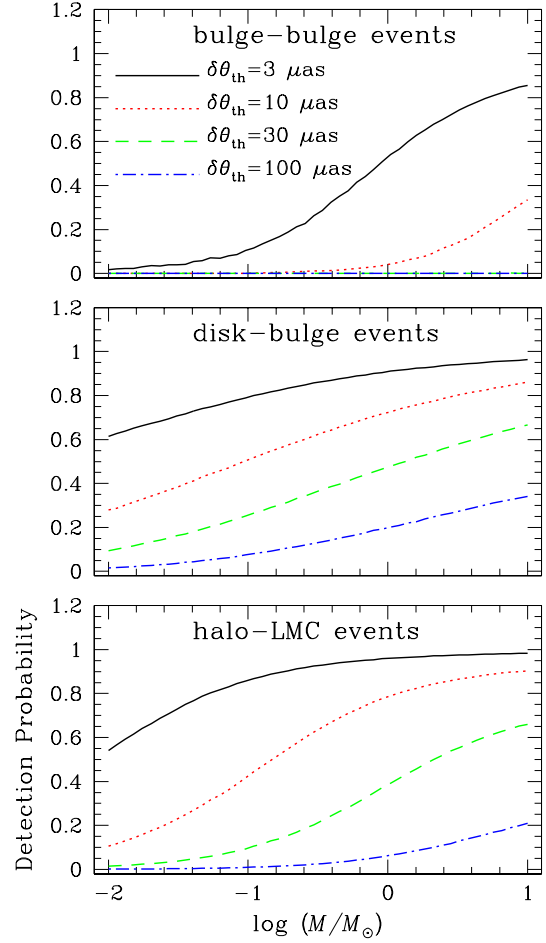}}
\caption{
The distributions of the probability of detecting the
parallax-induced deviations in the centroid shift trajectories for given
values of detection threshold, $\delta\theta_{\rm th}$, as a function of
(logarithmic) lens mass. 
}
\end{figure}

\section{Detection of Parallax-induced Deviations}

In the previous section, we show that the amount of the maximum centroid
shifts for most Galactic microlensing events are large enough to be detected
both from space and ground-based astrometric observations.  However, as shown
in the example of Figure 1, the amount of the parallax-induced deviations
will be much smaller than the shifts caused by the lensing itself.  Then
a naturally arising question is that for what fraction of events one can
determine lens parallaxes by detecting $\Deltavec\deltavec\thetavec_{\rm c}$.
In this section, we answer this question.

To determine the fraction of events with detectable $\Deltavec\deltavec
\thetavec_{\rm c}$, we compute the distribution of the average
maximum deviations by
\begin{equation}
\eqalign{
f(\langle\Delta\delta\theta_{\rm c,max}\rangle) & = \int_0^\infty dD_{os} 
\rho(D_{os}) \int_0^{D_{os}} dD_{ol} \rho(D_{ol})\pi r_{\rm E}^2 \cr
 & \hskip-55pt\times \int_0^\infty \int_0^\infty dv_x dv_y v f(v_x,v_y) \cr
 & \hskip-55pt\times \delta \left[ \langle\Delta\delta\theta_{\rm c,max}\rangle - 
   \langle\Delta\delta\theta_{\rm c,max}\rangle' 
   \left(\theta_{\rm E},\Pi; D_{ol},D_{os},M\right) \right].  \cr
}
\end{equation}
For the computation of $\langle\Delta\delta\theta_{\rm c,max}\rangle$,
we first compute $\Delta\delta\theta_{\rm c}$ of each event for given values
of the lens parameters ($M$, $D_{ol}$, and $D_{os}$) and the parameters of
the relative observer-lens-source positions ($\lambda_\odot$ and $\phi$) by
using the equations in \S\ 2.  We then obtain the maximum value which is
expected from the assumed duration of observation.  Since astrometric
observations will be carried out for events detected from photometric
monitoring, we assume observation duration to be $-1 t_{\rm E} \leq 
t_{\rm obs} \leq 10 t_{\rm E}$.  Since the lens proper motion vector $\muvec$
has no preferred orientation, the average value of the maximum centroid
shifts is obtained by assuming that $\phi$ is randomly distributed
in the range $0 \leq \phi \leq 2\pi$.  In addition, since we are interested
in all lensing events not in the events detectable by a specific lensing
survey, the impact parameters of events have random distribution in the
range $0 \leq u_{\rm min} \leq 1.0$.  The ecliptic latitude of a star towards
the Baade's window is $\beta\sim -6^\circ\hskip-2pt .5$.  For the physical
and dynamical distributions of the lens and source, we assume the same
distributions used for the computation of $f(\delta\theta_{\rm c,max})$.

Figure 3 shows the obtained distributions $f(\langle \Delta\delta
\theta_{\rm c,max}\rangle)$ for various types of Galactic lensing events.
In Figure 4, we also present the distributions of the probability of detecting
$\Deltavec\deltavec\thetavec_{\rm c}$ for given values of detection threshold,
$\delta\theta_{\rm th}$, as a function of (logarithmic) lens mass.  From the
figures, one finds the following facts.  First, if observations are performed
with the ground interferometers, it will be difficult to detect $\Deltavec
\deltavec \thetavec_{\rm c}$ for nearly all bulge self-lensing events since
the detection probability is less than 1\% even for events caused by lenses
with $M=1.0\ M_\odot$.  In addition, detection will be limited for only small
fractions of disk-bulge and halo-LMC events.  For example, for these types
of events caused by $M=0.3\ M_\odot$, the probabilities are $36\%$ and $21\%$,
respectively.  Second, if events are observed by using the SIM, on the other
hand, it is expected to detect the deviations for majority of disk-bulge and
halo-LMC events and even for some fraction of bulge self-lensing events.  For
example, the detection probabilities for events with $M=0.3\ M_\odot$ are
86\% and 93\% for disk-bulge and halo-LMC events and 27\% for bulge
self-lensing events.  In Table 1, we summarize the detection probabilities
which are expected from the observations by using the SIM and ground
interferometers.

\begin{table}
\smallskip
\begin{center}
\begin{tabular}{cccc}
\hline
\hline
\multicolumn{1}{c}{\ event\ \  } &
\multicolumn{1}{c}{lens mass} &
\multicolumn{2}{c}{detection probability} \\
\multicolumn{1}{c}{type} &
\multicolumn{1}{c}{($M_\odot$)} &
\multicolumn{1}{c}{\ \ \ SIM\ \ \ } &
\multicolumn{1}{c}{\ \ ground\ \ } \\
\hline
bulge-bulge & 0.1 & 0.11 & 0.00 \\
            & 0.3 & 0.27 & 0.00 \\
            & 0.5 & 0.37 & 0.00 \\
\smallskip
            & 1.0 & 0.53 & 0.00 \\

disk-bulge  & 0.1 & 0.79 & 0.26 \\
            & 0.3 & 0.86 & 0.36 \\
            & 0.5 & 0.88 & 0.41 \\
\smallskip
            & 1.0 & 0.91 & 0.47 \\

halo-LMC    & 0.1 & 0.86 & 0.10 \\
            & 0.3 & 0.93 & 0.21 \\
            & 0.5 & 0.94 & 0.28 \\
            & 1.0 & 0.96 & 0.39 \\
\hline
\end{tabular}
\end{center}
\caption{The detection probabilities of the parallax-induced deviations
in centroid shift trajectories which are expected from the astrometric
observations by using the SIM and ground interferometers.  The assumed
detection thresholds are $\delta\theta_{\rm th}=3\ \mu {\rm as}$ for the
SIM and $\delta\theta_{\rm th}=30\ \mu {\rm as}$ for the ground
interferometers.}
\end{table}

\section{Summary and Conclusion}

We compute the distributions of the maximum astrometric source star image
centroid shifts, $\delta\theta_{\rm c,max}$, and the average maximum
deviations in the centroid shift trajectories, $\langle\Delta\delta
\theta_{\rm c,max}\rangle$, expected for different types of Galactic events
caused by various masses.  The findings from the analysis of these
distributions are summarized as follows.
\begin{enumerate}
\item
As long as source stars are bright enough for astrometric observations,
one can detect $\deltavec\thetavec_{\rm c}$ for most events caused by lenses
with masses greater than $0.1\ M_\odot$ regardless of the event types from
observations by using not only the SIM but also the ground interferometers.

\item
If observations are performed with the ground interferometers, detecting
$\Deltavec\deltavec\thetavec_{\rm c}$ will be possible for nearly none of
bulge self-lensing events and for only a small fractions of disk-bulge and
halo-LMC events.

\item
However, if events are observed by using the SIM, detection of $\Deltavec
\deltavec\thetavec_{\rm c}$ will be possible for most disk-bulge and
halo-LMC events and even for some fraction of bulge self-lensing
events.
\end{enumerate}
Therefore, for the complete resolution of the lens degeneracy and accurate
lens mass determination, SIM observations (or equivalent) will be essential.

\section*{Acknowledgments}
This work was supported by the International Cooperation Research Fund from
the Korea Science and Engineering Foundation (KOSEF).

\end{document}